\documentclass[aps,
prl,
reprint,
amsmath,
amssymb,
superscriptaddress,
preprintnumbers]{revtex4-2}

\usepackage[utf8]{inputenc}
\usepackage{epsfig}
\usepackage{hyperref}
\usepackage{breakurl}
\usepackage{tensor}
\usepackage{bm}
\usepackage{color}
\usepackage{tikz}
\usepackage{subfigure}

\usetikzlibrary{decorations}
\usetikzlibrary{trees}
\usetikzlibrary{decorations.pathmorphing}
\usetikzlibrary{decorations.markings}
\usetikzlibrary{external}
\usetikzlibrary{intersections}
\usetikzlibrary{shapes,arrows}
\usetikzlibrary{arrows.meta}
\usetikzlibrary{calc}
\usetikzlibrary{shapes.misc}
\usetikzlibrary{decorations.text}
\usetikzlibrary{backgrounds}
\usetikzlibrary{tikzmark,calc,arrows,shapes,decorations.pathreplacing}

\tikzset{
	graviton/.style={decorate,line width=0.25mm, decoration={snake,amplitude=.5mm, segment length=2mm}},
	massive/.style={postaction={decorate},
		line width=0.4mm,
	},
}
 
\makeatletter
\def\simgt{\mathrel{\lower2.5pt\vbox{\lineskip=0pt\baselineskip=0pt
           \hbox{$>$}\hbox{$\sim$}}}}
\def\simlt{\mathrel{\lower2.5pt\vbox{\lineskip=0pt\baselineskip=0pt
           \hbox{$<$}\hbox{$\sim$}}}}
\makeatother

\def\eqn#1{Eq.~\eqref{#1}}
\def\eqns#1.#2{Eqs.~\eqref{#1} and~\eqref{#2}}
\def\spa#1.#2{\left\langle#1\,#2\right\rangle}
\def\spb#1.#2{\left[#1\,#2\right]}
\def\sand#1.#2.#3{%
\left\langle#1{\vphantom1}\right|{#2}\left|#3\right]}%
\def\sandmp#1.#2.#3{%
\left\langle#1{\vphantom1}\right|{#2}\left|#3\right]}%
\def\sandpm#1.#2.#3{%
\left[#1{\vphantom1}\right|{#2}\left|#3\right\rangle}%
\def\sandmm#1.#2.#3{%
\left\langle#1{\vphantom1}\right|{#2}\left|#3\right\rangle}%
\def\sandpp#1.#2.#3{%
\left[#1{\vphantom1}\right|{#2}\left|#3\right]}%

\def\LL{\mathcal{L}}
\def\phis{\phi_{s}}
\def\CC{C_{2}}
\def\PLS{\mathbb{S}}
\def\Es{\mathcal{E}}
\def\gS{\mathsf{S}}
\def\oS{\mathsf{S}}
\def\sS{\mathsf{S}}
\def\order{O}
\def\op{\mathcal{O}}
\def\opK{\mathcal{K}}

\def\hdelta{{\hat\delta}}

\def\opH{\mathcal{H}}
\def\opp{\bm p}
\def\opr{\bm r}
\def\opL{\left(\opr \times \opp\right)}
\newcommand{\spinStr}[1]{\Sigma_{#1}}
\def\clS{\textbf{S}}
\def\clK{\textbf{K}}
\def\KxS{X}
\def\clKxS{\textbf{\KxS}}
\def\eftSigma{\sigma}

\def\doe{\partial}
\def\bs{\boldsymbol}

\def\clp{\bm p}

\newcommand{\sym}[1]{\{#1\}}

\def\nn{\nonumber}

\newcommand{\Section}[1]{\section{#1}}
\newcommand{\be}{\begin{equation}}
\newcommand{\ee}{\end{equation}}

\renewcommand{\imath}{\mathrm{i}}

\def\topbotatom#1{\hbox{\hbox to 0pt{$#1\bot$\hss}$#1\top$}}

\begin{document}

\title{Conservative Spin Magnitude Change in Orbital Evolution in General Relativity}

\author{Mark Alaverdian}
\affiliation{Institute for Gravitation and the Cosmos,
	Pennsylvania State University,
	University Park, PA 16802, USA}
\author{Zvi~Bern}
\affiliation{
	Mani L. Bhaumik Institute for Theoretical Physics,
	University of California at Los Angeles,
	Los Angeles, CA 90095, USA}
\author{Dimitrios~Kosmopoulos}
\affiliation{
         D\'epartement de Physique Th\'eorique, 
         Universit\'e de Gen\`eve, 
         CH-1211 Geneva, 
         Switzerland
}		
\author{Andres~Luna}
\affiliation{
	Niels Bohr International Academy,
	Niels Bohr Institute, University of Copenhagen,
	Blegdamsvej 17, DK-2100, Copenhagen \O , Denmark}
\author{Radu~Roiban}
\affiliation{Institute for Gravitation and the Cosmos,
	Pennsylvania State University,
	University Park, PA 16802, USA}
\author{Trevor~Scheopner}	
\affiliation{
	Mani L. Bhaumik Institute for Theoretical Physics,
	University of California at Los Angeles,
	Los Angeles, CA 90095, USA}
\author{Fei~Teng}
\affiliation{Institute for Gravitation and the Cosmos,
	Pennsylvania State University,
	University Park, PA 16802, USA}

\begin{abstract}

We show that physical scattering observables for compact spinning objects in general relativity can depend on additional degrees of freedom in the spin tensor beyond those described by the spin vector alone.  
The impulse, spin kick, and leading-order waveforms 
exhibit such a nontrivial dependence.
A signal of this additional structure is the change in the magnitude of the spin vector under conservative Hamiltonian evolution, similar to our previous studies in electrodynamics.
These additional degrees of freedom describe dynamical mass multipoles of compact objects and decouple for black holes. 
We also show that the conservative impulse, spin kick and change of the additional degrees of freedom are encoded in the eikonal phase.

\end{abstract}
   
\maketitle

\Section{Introduction}
The detection of gravitational waves by the LIGO/Virgo collaboration~\cite{LIGOScientific:2016aoc, LIGOScientific:2017vwq} has opened a new era in astronomy, cosmology, and perhaps even particle physics. As the sensitivity of gravitational-wave detectors will continue to improve~\cite{Punturo:2010zz, LISA:2017pwj, Reitze:2019iox}, the spin of compact astrophysical objects and the rich three-dimensional dynamics it introduces will become increasingly important for signal identification and interpretation. 
For further details, see reviews~\cite{Blanchet:2006zz, Porto:2016pyg, Levi:2018nxp, Buonanno:2022pgc} and references therein. 

The study of spinning compact objects with amplitudes methods, e.g.~\cite{Arkani-Hamed:2017jhn, Guevara:2018wpp, Chung:2018kqs, Guevara:2019fsj, Maybee:2019jus, Chung:2019duq,  Damgaard:2019lfh, Bern:2020buy, Kosmopoulos:2021zoq, Aoude:2022thd, Cangemi:2022bew, Aoude:2023vdk, Cangemi:2023ysz, Cangemi:2023bpe},
brought to the forefront interesting and subtle theoretical questions regarding the number of physical degrees of freedom and independent operators describing their interactions~\cite{Bern:2022kto, Bern:2023ity}.
Here, we discuss these issues in general relativity 
following Ref.~\cite{Bern:2023ity}, which studied them in electrodynamics.

In traditional worldline approaches~(see, e.g., Refs.~\cite{Porto:2005ac,  Steinhoff:2015ksa, Levi:2015msa, Liu:2021zxr}) as well as in world-line quantum field theory~\cite{Jakobsen:2021zvh}, a spin supplementary condition (SSC)~\cite{Fleming} identifies the three physical spin degrees of freedom and restricts their interactions~\footnote{The SSC was interpreted in Ref.~\cite{Steinhoff:2015ksa, Levi:2015msa, Vines:2016unv} in terms of a spin-gauge symmetry encoding the freedom to locally shift the worldline in the ambient space.}.
While a consequence of the SSC is the invariance of the spin magnitude under conservative time evolution, rotation and Lorentz invariance generally guarantee only the conservation of the (magnitude of the) total angular momentum.

Suitably relaxing the SSC introduces new dynamical degrees of freedom and additional interactions tagged by additional Wilson coefficients. Ref.~\cite{Bern:2023ity} studied the effective electromagnetic interactions of such spinning particles from field theory, worldline, and two-body Hamiltonian perspectives and found all approaches consistent.
In field theory, the new degrees of freedom relate to the description of 
the classical spin as a superposition of quantum spins with allowed 
transitions. 
On the worldline, they are components of the spin tensor,
\begin{equation}\label{eq:gStensor}
\sS^{\mu\nu} = \frac{1}{m}\epsilon^{\mu\nu\rho\sigma}p_\rho S_\sigma + \frac{i}{m}(p^{\mu} K^{\nu}-p^{\nu}K^{\mu})\, .
\end{equation}
The new dynamical variable $K^\mu$ has been interpreted as a ``mass dipole"~\cite{Costa:2011zn, Costa:2014nta} and $S^\mu$ is the usual spin vector. They satisfy $p^{\mu}S_{\mu}=p^{\mu}K_{\mu}=0$.
Finally, the two-body Hamiltonian must include operators $\clS$ and $\clK$~\cite{Bern:2023ity} to reproduce the field theory and worldline results with the extra degrees of freedom.
In gravity, Refs.~\cite{dAmbrosi:2015ndl, vanHolten:2015vfa, dAmbrosi:2015wqz} understood the self-consistency of the worldline approach with a dynamical mass dipole and no SSC, and explored its physical inequivalence.

Here, we show that the conclusions of Ref.~\cite{Bern:2023ity} for electrodynamics carry over to gravity, including the spin-magnitude evolution under conservative Hamiltonian dynamics.
While the linear-in-spin (dipole) interactions are not altered by the presence of additional degrees of freedom, in accordance with the principles of general relativity, physical observables such as scattering angles and waveforms reflect their presence starting at the second order in the spin tensor \eqref{eq:gStensor}~\footnote{This contrasts with electrodynamics, where the first effect is linear in the spin tensor, and reflects the lack of universality of electromagnetic interactions at linear order in spin.}.
Starting at $\order(G^2 S^3)$, artificial exclusion of the extra degrees of freedom in the external states does not remove the additional Wilson coefficients from observables~\cite{Bern:2022kto}. 
Interestingly, for special values of Wilson coefficients, the dynamics of $K$ decouples from the spin and orbital dynamics, indicating the emergence of a local symmetry, presumably related to spin-gauge symmetry~\cite{Levi:2015msa}. This interpretation is compatible with the Kerr stress tensor~\cite{Vines:2017hyw}, the results of Teukolsky equations~\cite{Saketh:2022wap, Bautista:2022wjf, Bautista:2023szu} and with earlier results for black-hole scattering through $\order(G^2 S^4)$~\cite{Bern:2020buy, 
Liu:2021zxr, Kosmopoulos:2021zoq, Chen:2021kxt, Aoude:2022trd, Bern:2022kto}~\vphantom{\cite{Saketh:2022wap, Bautista:2022wjf, Bautista:2023szu, Cangemi:2023bpe, Bautista:2023sdf}}\footnote{Through $\order(S^4)$ Refs.~\cite{Saketh:2022wap, Bautista:2022wjf, Bautista:2023szu} find that the black hole Compton amplitude obtained via the Teukolsky equation agrees with these previous results, but beyond this, the situation is less clear. See Refs.~\cite{Cangemi:2023bpe, Bautista:2023sdf} for recent discussions on reconciling the approaches based on scattering amplitudes and the Teukolsky equation.}.


\Section{Field Theory}

We use the amplitudes-based field-theory approach of Refs.~\cite{Bern:2020buy, Kosmopoulos:2021zoq, Bern:2022kto, Bern:2023ity}, and describe a spinning body by a symmetric traceless tensor field $\phis$. 
%
%
In the local frame, 
the minimal interaction with Einstein's gravity is,
\begin{align}\label{eq:Lmin}
    \LL_{0} &= - \frac{1}{2}\phis(\nabla^2+m^2)\phis + \frac{H}{8}R_{abcd}\phis M^{ab}M^{cd}\phis \,,
\end{align}
where $M^{ab}$ are the Lorentz generators acting in the space of $\phis$.
See, e.g., Ref.~\cite{Holstein:2008sx, Vaidya:2014kza,  Arkani-Hamed:2017jhn, Damgaard:2019lfh, Aoude:2022thd, FebresCordero:2022jts, Aoude:2023vdk, Cangemi:2023bpe} for fixed-spin field-theory approaches to spinning bodies.
We can include nonminimal interactions starting at $\order(M_{ab}^2)$,
\begin{align}
\label{eq:ExtraInteractions}
    &\LL_{\text{non-min}} = - \frac{\CC}{2m^2} R_{a f_1 b f_2}\nabla^a\phis \PLS^{(f_1}\PLS^{f_2)}\nabla^b\phis \nonumber\\
    &\quad + \frac{D_2}{2m^2}R_{abcd} \nabla_{i}\phis\sym{M^{ai}M^{cd}}\nabla^b\phis\\
    & \quad + \frac{E_2-2D_2}{2m^4}R_{abcd}\nabla^{(a}\nabla^{i)}\phis\sym{M\indices{^b_i}M\indices{^d_j}}\nabla^{(c}\nabla^{j)}\phis\,,\nonumber
\end{align}
where $\PLS^{a}={-i}\epsilon^{abcd}M_{bc}\nabla_{d}/(2m)$ and $\{\dots\}$ denotes symmetrization of the argument.  The $C_2$ term corresponds to the only interaction allowed on the traditional worldline. 
In contrast, in the field-theory formalism, it is natural to include all interactions pertinent to 
classical physics. The Wilson coefficients $(H, C_2, D_2, E_2)$ are
interdependent, \emph{allowing us to fix $H=1$.} 
Also, we define $E_2$ and $D_2$ to streamline the amplitudes and observables.

As Ref.~\cite{Bern:2023ity} discussed in some detail in electrodynamics through quadratic order in spin, demanding that only states of fixed spin propagate in diagrams enforces the conservation of the magnitude of the spin vector and is equivalent to enforcing the SSC. 
In contrast, when transitions between states of different spins are allowed, all interactions in \eqn{eq:ExtraInteractions} contribute to physical observables.  
While negative-norm states can appear in the electrodynamics analog of 
\eqn{eq:Lmin}, a more involved Lagrangian with only positive-norm states does not change this conclusion~\cite{Bern:2023ity}. Furthermore, physical observables obtained in the two cases are related by a simple mapping of parameters. 
Since the negative-norm states do not change the classical limit, we 
find it more practical to use \eqn{eq:ExtraInteractions}.

We choose boosted spin-coherent states as external polarizations $\Es_i$ of massive spinning particles and define the classical spin tensor $\gS^{ab}$ as the expectation value of Lorentz generators $M^{ab}$ in these states (see Eq.~(2.27) of Ref.~\cite{Bern:2023ity}).
Due to the presence of degrees of freedom beyond those of a single fixed spin, including longitudinal components in the external states, 
the product of massive polarization tensors also depends on the $K$ vector~\cite{Bern:2023ity},
\begin{align}\label{eq:EE2}
    \Es_1\cdot\Es_2 = \exp\left[\frac{q\cdot K}{m}\right]\Es_1^{(s)}\cdot\Es_2^{(s)} + \order(q)\,,
\end{align}
where $q=p_2-p_1$, $p_1$ and $p_2$ are incoming and outgoing momenta of a particle of mass $m$ and $\Es^{(s)}_i$ is the transverse traceless component of $\Es_i$ and corresponds to the coherent state of a fixed spin $s$.
The product $\Es_1^{(s)}\cdot\Es_2^{(s)}$ only depends on the spin vector and connects the covariant and canonical impact 
parameter~\cite{Bern:2020buy}. 

We construct the three- and four-point classical 
amplitudes with two $\phi_s$ fields and gravitons
using the Feynman rules from the Lagrangian $\LL_0+\LL_{\text{non-min}}$. 
We use Eq.~\eqref{eq:gStensor}  to separate $S^\mu$ and $K^\mu$,
and expose the complete $K$-dependence, but suppress the spin-vector-dependent 
factor $\Es_1^{(s)}\cdot\Es_2^{(s)}$ 
and factors of $\kappa$ with $\kappa^2 = 32 \pi G$.
With graviton polarizations written as $\varepsilon_i^{\mu\nu}=\varepsilon_i^\mu\varepsilon_i^\nu$, these amplitudes through $\order(\oS^2)$ are
\begin{widetext}
\begin{align}\label{eq:MComp}
\mathcal{M}^{\text{3pt}} &= -(\varepsilon_1\cdot p)^2 + \frac{(\varepsilon_1\cdot p)\tilde{f}_1(p,S)}{m} - \frac{(1+C_2)(\varepsilon_1\cdot p)^2(k_1\cdot S)^2}{2m^2} -\frac{D_2(k_1\cdot K)(\varepsilon_1\cdot p)\tilde{f}_1(p,S)}{m^2} - \frac{E_2(k_1\cdot K)^2(\varepsilon_1\cdot p)^2}{2m^2} \,,\nonumber\\
\mathcal{M}^{\text{4pt}}_\text{C} &= \frac{4}{\hat{s}{t}\hat{u}}\left[\alpha^2 - \alpha\,\op_{(1)} + \frac{1}{2}\op_{(1)}^2 + \CC \, \alpha \, \op_{(2)}^{} + D_2 \, \alpha \left(\op_{(1)}^{}\frac{(k_{1}+k_2)\cdot K}{m} - \opK_{(1,1)}^{}\right) + E_2 \left(\alpha\, \op_{(2)}\Big|_{S\rightarrow K}\right)\right]\, .
\end{align}
\end{widetext}
Here $\alpha = p \cdot f_1 \cdot f_2 \cdot p$, and $f_i^{\mu\nu} = k_i^{\mu}\varepsilon_i^{\nu} -k_i^{\nu}\varepsilon_i^{\mu}$ is the linearized field strength.
The kinematic variables are defined as $\hat{s}=2p\cdot k_1$, ${t}=2k_1\cdot k_2$ and $\hat{u}=2p\cdot k_2$, 
while $\op_i$ and $\opK$ are given by
\begin{align}
    & \op_{(1)} = \frac{1}{m}\Big[f_2(p,k_1)\tilde{f}_1(p,S)+\frac{\hat{s}}{2}\,\tilde{f}_{12}(p,S) +(1\leftrightarrow 2)\Big]\,, \nonumber\\
    & \op_{(2)} = \frac{1}{2m^2} \Big[ t f_1(p_,S)f_2(p,S)+\alpha (k_{1}\cdot S+k_2\cdot S)^2\Big]\,,\nonumber\\
    & \opK_{(1,1)} = \frac{t}{2m^2}\Big[f_2(p,K)\tilde{f}_1(p,S)+f_1(p,K)\tilde{f}_2(p,S)\Big]\,,
\label{eq:Operators}
\end{align}
where $f_i(a,b) = f_i^{\mu\nu}a_{\mu}b_{\nu}$, and $(\tilde{f}_i^{\mu\nu},\tilde{f}_{12}^{\mu\nu})$ are the Hodge duals~\footnote{We define the Hodge dual as $\tilde{f}^{\mu\nu}=\frac{i}{2}\epsilon^{\mu\nu\alpha\beta}f_{\alpha\beta}$} of $(f_i^{\mu\nu},f^{\mu\rho}_1f\indices{_{2,\rho}^{\nu}})$. 

Several comments are in order. In line with the absence of interactions of mass dipoles in general relativity and despite the presence of $K$ encoding the longitudinal components of the external states, the $\order(\oS)$ terms are unchanged as there are no linear-in-$K$ contributions at this order.
At $\order(\oS^2)$, $K$ enters schematically as $SK$ and $K^2$. Interestingly, the $K^2$ dependence is identical to the $S^2$ dependence with Wilson coefficient $C_2$. The amplitudes in \eqn{eq:MComp} are independent of $K$ if $D_2=E_2=0$, as the $K$ vectors originating from interactions exactly cancel the contribution of the exponential factor in Eq.~\eqref{eq:EE2}.  
As noted below, for these values of the Wilson coefficients, our results agree with the worldline formalism of Ref.~\cite{Levi:2015msa}, which enforces the SSC. 
Finally, the Kerr black hole corresponds to further setting $C_2=0$, in close analogy with the root-Kerr solution in electrodynamics~\cite{Bern:2023ity}.

Starting from the amplitudes in Eq.~\eqref{eq:MComp}, we compute two-body scattering amplitudes at $\order(G)$ and $\order(G^2)$ using generalized unitarity~\cite{Bern:1994zx, Bern:1994cg, Kosmopoulos:2020pcd}, 
from which we may obtain the classical two-body effective Hamiltonian and observables through a matching process~\cite{Cheung:2018wkq, Bern:2020buy}. 
These amplitudes are collected in the ancillary file {\tt Amplitudes.m}.


\Section{Worldline}
In the standard worldline approaches to spinning particles, an SSC such as $p_\mu \sS^{\mu\nu} = 0$ corresponds to a symmetry~\cite{Steinhoff:2015ksa, Vines:2016unv, Jakobsen:2021zvh}. This eliminates $K^\mu$ in \eqn{eq:gStensor} as a dynamical variable,  forbids certain interaction terms, and leads to spin-vector magnitude conservation.
High-order calculations in spin or coupling (see e.g., Refs.~\cite{Levi:2019kgk, Siemonsen:2019dsu, Levi:2020kvb, Levi:2020uwu, Levi:2020lfn, Liu:2021zxr, Cho:2021mqw, Kim:2021rfj, Jakobsen:2022fcj, Cho:2022syn, Kim:2022pou, Kim:2022bwv, Levi:2022dqm, Levi:2022rrq, Jakobsen:2023ndj, Jakobsen:2023hig, Ben-Shahar:2023djm, Scheopner:2023rzp}) have been carried out in such a manner. 

Ref.~\cite{Bern:2023ity} discussed a worldline theory with no SSC, rediscovering the observation of Refs.~\cite{dAmbrosi:2015ndl, vanHolten:2015vfa, dAmbrosi:2015wqz} that such theories are self-consistent and can be distinct. 
Here, we construct the analogous gravitational worldline theory and show that its Compton amplitude matches that of the abovementioned field theory. 

The dynamical variables are a timelike worldline $z^\mu(\lambda)$, its conjugate momentum $p_\mu(\lambda)$, a body tetrad $\Lambda^A{}_\mu(\lambda)$, and the body spin tensor $\sS^{\mu\nu}(\lambda)$, where $\lambda$ is the worldline affine parameter. The Lagrangian is
\begin{equation}\label{eq:wlact}
L = -p_\mu \dot z^\mu + \frac{1}{2} \sS^{\mu\nu}\Lambda_{A\mu} \frac{D\Lambda^A{}_{\nu}}{D\lambda} + \frac{\xi}{2}(p^2 - \text{M}^2)\,,
\end{equation}
where $\xi(\lambda)$ is a Lagrange multiplier that enforces the on-shell constraint and $\text{M}\equiv \text{M}(z,p,\sS)$ is the dynamical mass function of the body~\cite{JanSteinhoff:2015ist}, which contains the body's free mass and all its non-minimal couplings to gravity. 
Extremization of the action $\int d\lambda L$ yields the body's dynamics.
In terms of ${\hat p}{}^\mu = {p^\mu}/{\sqrt{p^2}}$ and to second order in the vectors $S^\mu$ and $\text{K}^\mu = i K^\mu$~\footnote{The redefinition $\text{K}^\mu = i K^\mu $ addresses the anti-hermiticity of the $K$ vector in Eq.~\eqref{eq:gStensor}. The variable $\text{K}^\mu$ appears only in our worldline discussion and its rest frame form $\clK$ appears in the Hamiltonian. The waveform discussion below uses $K^\mu$ and its rest frame form $\bm K$.}, the dynamical mass function can be written as:
\begin{align}
\label{eq:dmf}
    \text{M}^2 &= m^2 \\
    &+ \frac{1+C_2}{4}R_{{\hat p} S {\hat p} S} + \frac{1+D_2}{2} R^\star_{{\hat p} S {\hat p} \text{K}} + \frac{1+E_2}{4} R_{{\hat p} \text{K} {\hat p} \text{K}} \,,  \nn 
\end{align}
with $R_{{\hat p} S {\hat p} S} = R_{\mu\nu\rho\sigma}{\hat p}{}^\mu S^\nu {\hat p}{}^\rho S^\sigma$ and $R^\star_{\mu\nu\rho\sigma}$ is the right-dual of the Riemann tensor. $\text{K}$ is defined in terms of the spin tensor as $\text{K}^{\mu} = - \mathsf{S}^{\mu\nu}\hat{p}_{\nu}$. In general, the map between 
the worldline and QFT Wilson coefficients can be nontrivial. Eq.~\eqref{eq:dmf} implements this map at ${\cal O}(S^2)$, aligning Wilson coefficients.

The classical gravitational Compton amplitude ${\cal M}_\text{C}^{\mu\nu, \rho\sigma}$ is computed \cite{Saketh:2022wap} as the coefficient of the outgoing spherical wave produced by the response of the spinning body to an incoming plane wave. The metric perturbation and the amplitude are extracted directly from the stress tensor $T_{\mu\nu} = \frac{\delta}{\delta h^{\mu\nu}}\int d\lambda L$ using the solution to the wave equation at large distances in de Donder gauge.   
${\cal M}_\text{C}^{\mu\nu, \rho\sigma}$
reproduces the second line of \eqn{eq:MComp}, pointing to the equivalence of this worldline theory and the field theory described above.

Mirroring the discussion of the field-theory Compton amplitudes and similarly to the case of electrodynamics~\cite{Bern:2023ity}, for the special values of the additional Wilson coefficients, $D_2=E_2=0$, 
the SSC condition $p_\mu \sS^{\mu\nu} = 0$, i.e. $\text{K}^\mu=0$, is preserved under time evolution.
%
%
Thus, from a practical standpoint, if an SSC is desired, it is convenient to impose it by simply choosing Wilson coefficients that decouple the dynamics of the $K$ vector.

\begin{figure*}[tb]
        \includegraphics[height=1.82in]{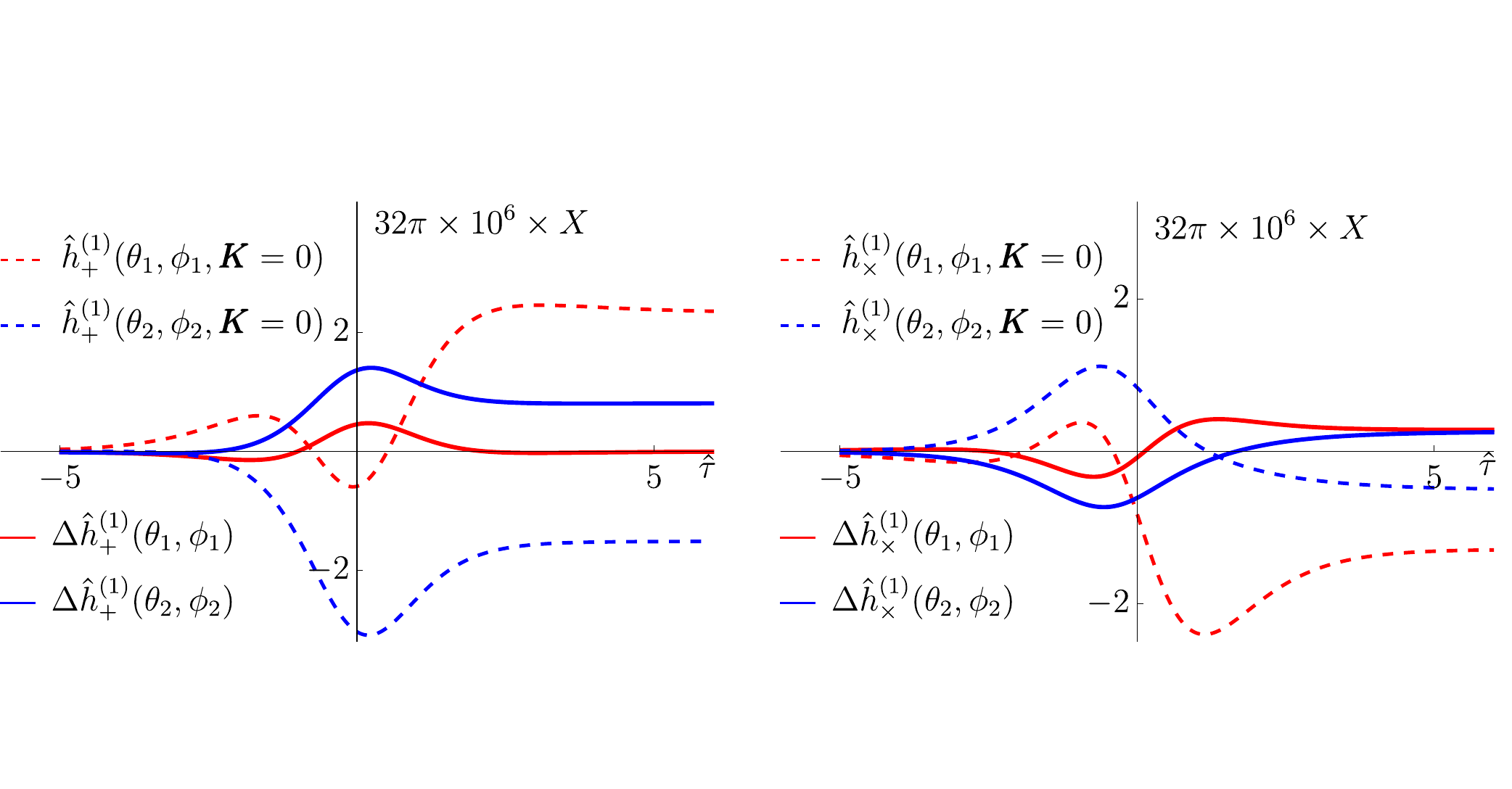}
        \includegraphics[height=1.82in]{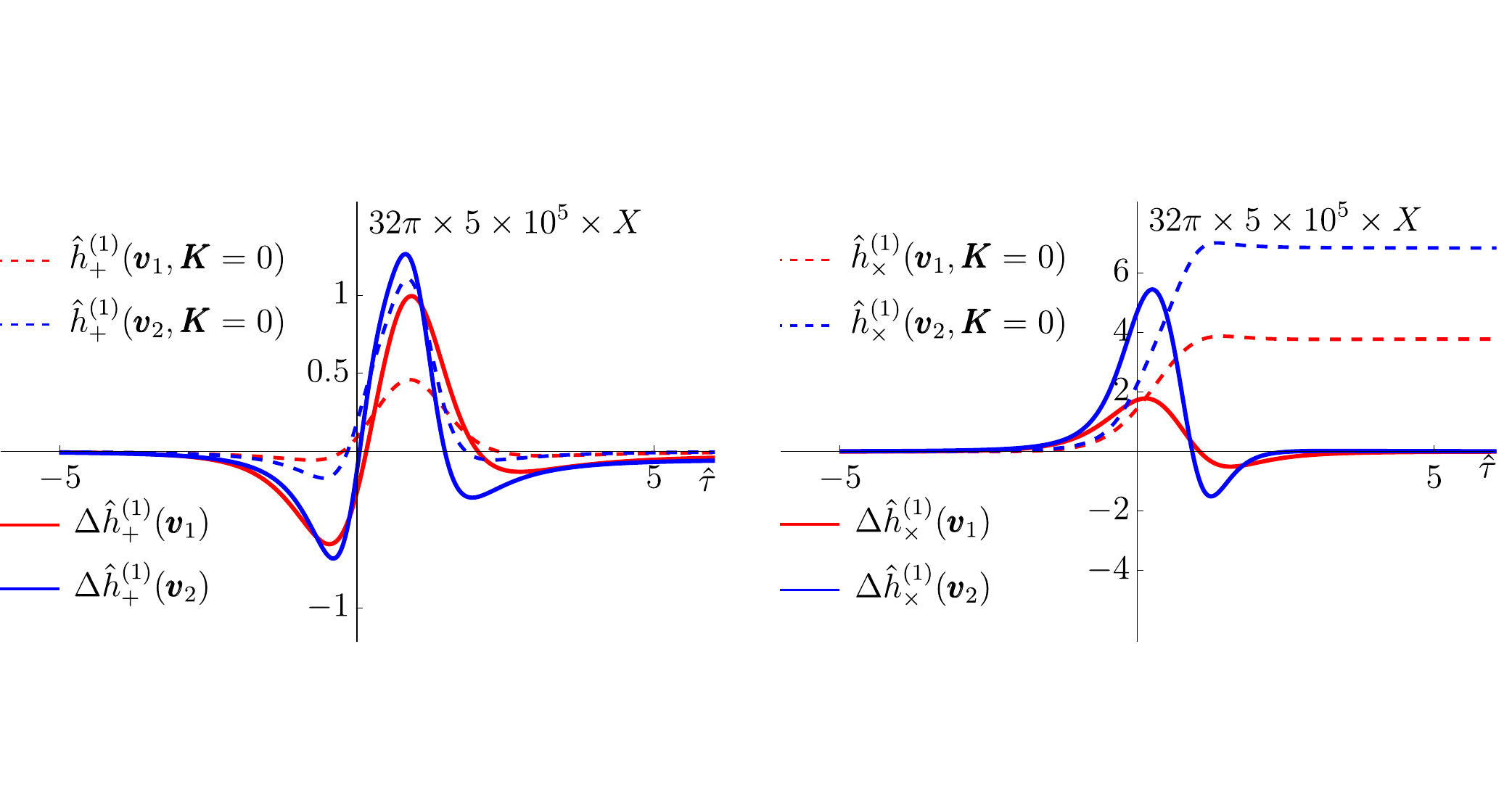}
 \vskip -.5 cm 
\caption{
{\em Dashed lines:} Rescaled leading order numerical scattering waveforms, $\varepsilon^{\mu\nu} {\hat h}^{(1)}_{\mu\nu}=|b|W^\text{LO}/(\kappa^3 (m_1+m_2)^2)$, for equal-mass particles, $m_1=m_2 = m$,  and $\bm S = 
(1,1, 0)/\sqrt{2}$, $\bm K = \bm 0$ and $C_2 = 1$, denoted by $h_{+/\times}^{(1)}(\dots, \bm K = 0)$.
{\em Solid lines:} The differences $\Delta h_{+/\times}^{(1)} = h_{+/\times}^{(1)}(\dots, \bm K = \bm S)-h_{+/\times}^{(1)}(\dots, \bm K = 0)$; the former has $C_2\ne 1$, corresponding to an $E_2\ne 0$.
The parameters are as follows. {\em Left two figures:} wave sourced by particles with COM velocity $\bm v = (0, 0, 1/5)$ observed at angles
$(\theta_1, \phi_1)=(7, 4)\pi/10$ and $(\theta_2, \phi_2)=(1, 4)\pi/10$. Here $C_2=1/3$ and $D_2\simeq 0.287$ 
is chosen so that 
$\Delta \hat h_+^{(1)}(\theta_1, \phi_1)|_{\tau\rightarrow +\infty}=0$ at velocity $\bm v$.
{\em Right two figures:} waves sourced by particles with COM velocities 
$\bm v_1=(0, 0, 1/3)$, $\bm v_2=(0, 0, 1/2)$
observed at angle 
$(\theta, \phi) = (5, 10)\pi/10$. Here $C_2=99/100$ and
$D_2\simeq 2.421$ 
is chosen so that
$\Delta \hat h_\times^{(1)}(\bm v_1)|_{\tau\rightarrow +\infty}=0$  
at the observation angle.
We use the covariant impact parameter defined in Ref.~\cite{Bern:2020buy} and we choose it to be 
$m\, b_{\text{cov}}^{\mu}=(0,m \, \bm b_{\text{cov}}) = (0, 50, 0, 0)$ 
in units of the particles' mass. 
Finally, the vertical axis label $X$ is given in the legend of each plot, and $\hat\tau=\tau/|\bm b_{\text{cov}}|$. 
}
\label{fig:waveforms}
\end{figure*}

\Section{Scattering waveform at leading order}
To explore the physical relevance of the $K$ vector and whether observations can distinguish it from the spin,  we assume that we are given a waveform signal that can be fitted by a ${K}^\mu=0$ system and study whether it is possible to accurately describe the same signal by turning on $K$ and readjusting Wilson coefficients.
Since our spin-dependent amplitudes are given to a fixed (second) order in spin, we compare each order separately. 

Because $K^\mu$ does not enter at $\order(\sS)$, $S^\mu$ should remain fixed as $K$ is turned on. Thus, we may only adjust the Wilson coefficients. 
We find that to the order we are working and unless $D_2=0$, the waveform at a fixed observation angle discerns a nonvanishing $K$ vector~\vphantom{\cite{DeAngelis:2023lvf, Brandhuber:2023hhl, Aoude:2023dui, Jakobsen:2021lvp}}\footnote{The leading-order waveform for the scattering of a Kerr black hole off a spinless body was computed to all orders in spin in Refs.~\cite{DeAngelis:2023lvf, Brandhuber:2023hhl, Aoude:2023dui}. We have verified that our $K=0$ results exactly agree with Ref.~\cite{DeAngelis:2023lvf}.  Leading order waveforms through $\order(S^2)$ were also computed from the worldline in Ref.~\cite{Jakobsen:2021lvp}}.

Ref.~\cite{Cristofoli:2021vyo} established the connection between the time-dependent metric on the celestial sphere and scattering amplitudes. At leading order it takes the simple form
\begin{align}\label{eq:MtoWEFT}
W^{\text{LO}}(t) &= \frac{1}{\kappa}\lim_{|\bm x|\rightarrow\infty} 4\pi|\bm x| (g_{\mu\nu}-\eta_{\mu\nu}) \varepsilon^{\mu\nu}\Big|^\text{LO} \nonumber\\
&= - 2 \int_{-\infty}^{+\infty}\frac{d\omega}{2\pi}\int d\mu\,\mathcal{M}_{5}^{\text{tree}}e^{-i q_1\cdot b}e^{-i\omega t}\,,
\end{align}
where 
${\cal M}^\text{tree}_5\equiv {\cal M}^\text{tree}_5(p_1, p_2, q_1, k, \varepsilon)$ is the five-point amplitude with an on-shell outgoing graviton of momentum $k$ and polarization $\varepsilon$. The measure $d\mu$ is
\begin{align}\label{eq:cl_measure}
    d\mu = \frac{d^4 q_1}{(2\pi)^4} \hdelta(2p_1\cdot q_1)\hdelta\big(2p_2\cdot (k-q_1)\big)\,,
\end{align}
with $\hdelta(x) = 2\pi\delta (x)$, 
%
and $q_{1}$ is the momentum lost by particle $p_1$.

We focus on one spinning and one spinless particle. Using generalized unitarity, we obtain the tree-level classical five-point amplitude necessary for ${W}^\text{LO}$ by sewing together the two-matter, one- and two-graviton
(Compton) amplitudes.
Schematically, it is given by
\begin{align}
\pgfmathsetmacro{\w}{1.4}
\pgfmathsetmacro{\h}{1.3}
\pgfmathsetmacro{\x}{0.3}
\pgfmathsetmacro{\r}{0.17}
\mathcal{M}_5 \sim 
\vcenter{\hbox{\begin{tikzpicture}[scale=.8]
    \draw [massive] (0,0) node [left]{\small $1$} -- (\w,0);
    \draw [massive] (0,\h) node [left]{\small $2$} -- (\w,\h);
    \draw [graviton] (\w/2,0) -- ++ (0,\h);
    \draw [graviton] (\w/2,0) -- (\w,\h/2 - 0.1) node[above]{\small $k$};
    \draw [dashed] (\x,\h/2) -- ({\w-\x},\h/2);
    \filldraw [fill=gray!50] (\w/2,0) circle (\r);
    \filldraw [fill=gray!50] (\w/2,\h) circle (\r);
\end{tikzpicture}}}
+
\vcenter{\hbox{\begin{tikzpicture}[scale=.8]
    \draw [massive] (0,0) node [left]{\small $1$} -- (\w,0);
    \draw [massive] (0,\h) node [left]{\small $2$} -- (\w,\h);
    \draw [graviton] (\w/2,0) -- ++ (0,\h);
    \draw [graviton] (\w/2,\h) -- (\w,\h/2 + 0.1) node[below=0]{\small $k$};
    \draw [dashed] (\x,\h/2) -- ({\w-\x},\h/2);
    \filldraw [fill=gray!50] (\w/2,0) circle (\r);
    \filldraw [fill=gray!50] (\w/2,\h) circle (\r);
\end{tikzpicture}}}
-
\vcenter{\hbox{\begin{tikzpicture}[scale=.8]
    \draw [massive] (0,0) node[left]{\small $1$}-- (\w,0);
    \draw [massive] (0,\h) node[left]{\small $2$} -- (\w,\h);
    \draw [graviton] (\w/2,0) -- ++ (0,\h);
    \draw [graviton] (\w/2,\h/2) -- (\w,\h/2) node[right]{\small $\!k$};
    \draw [dashed] (\x,\h/4) -- ({\w-\x},\h/4);
    \draw [dashed] (\x,3*\h/4) -- ({\w-\x},3*\h/4);
    \filldraw [fill=gray!50] (\w/2,0) circle (\r);
    \filldraw [fill=gray!50] (\w/2,\h) circle (\r);
    \filldraw [fill=gray!50] (\w/2,\h/2) circle (\r);
\end{tikzpicture}}}\,,
\label{eq:sewing}
\end{align}
where the dashed lines indicate on-shell conditions and summation over physical states. We ignore the local terms as they do not contribute in the classical regime.

The gravitational-wave memory, given by the soft limit of ${\cal M}_5$, provides clues into the structure of the complete waveform. 
Interestingly, we find that it follows the same pattern as for the scattering of spinless particles~\cite{Saha:2019tub, Sahoo:2021ctw}, 
\begin{align}
\varepsilon^{\mu\nu}h^{\infty}_{\mu\nu}(\omega) \stackrel{\omega\rightarrow 0}{=}\frac{i}{\omega}\sum_{i=1}^{4}\frac{\eta_i(\varepsilon\cdot p_i)^2}{n\cdot p_i} + \order(\omega^{0}) \,,
\end{align}
where $k^{\mu}=\omega n^{\mu}$, $\eta_{1,2}=1$, $\eta_{3,4}=-1$ and $\varepsilon^{\mu\nu}=\varepsilon^{\mu}\varepsilon^{\nu}$. Momentum conservation requires $p_4 = p_1+\Delta p$ and $p_3 = p_2-\Delta p$, where $\Delta p$ is the leading-order covariant impulse included in the ancillary file~{\tt LOImpulse.m}. 

It is not difficult to see that the third term in Eq.~\eqref{eq:sewing} is subleading as $k^\mu \rightarrow 0$. Thus, the soft limit of the tree-level Compton amplitude determines the soft limit of the classical five-point tree amplitude.
In the former, the $\order(C_2S^2)$ terms are the same (up to $S\rightarrow K$) as the $\order(E_2 K^2)$ terms, see Eq.~\eqref{eq:Operators}.
It turns out that the same holds for the entire five-point amplitude.
That is, if we turn on $K^\mu\propto S^\mu$, we can compensate for the $K^2$ terms
by adjusting $C_2$.
Thus, if $D_2=0$, such a $K$ vector is degenerate with the spin.
In the following, we consider only $D_2\neq 0$ and $K^\mu \propto S^\mu$, which endows the system with an axial symmetry. 
To streamline the comparison, we interpret $E_2\ne 0$ as a change in $C_2$.

The memory contribution of the $\order(SK)$ terms indicates that we can choose the Wilson coefficients $E_2$ and $D_2$ so that $K$ is not distinguishable {\em at late times} in a fixed observation direction,
a fixed center-of-mass (COM) velocity $\bm v$ and one, say $+$, polarization.
At finite times, however, and at all times for the second polarization or for other velocities or observation directions, the difference between waveforms with and without $K$ is nontrivial, as illustrated for several configurations of velocities and observation angles in Fig.~\ref{fig:waveforms}.
The differences $\Delta h = W(K=S)-W(K=0)$ at fixed $S^\mu$ can be as large as about $50\%$ of the $K$-independent waveforms for $S$ parallel with $K$ and of comparable magnitude.
We note that the peak of the waveform difference for the $+$ polarization appears to exhibit a weak velocity dependence at fixed angles. 
Also, waveform differences are broader at lower velocities because, at fixed impact parameter, such particles experience a longer period of stronger acceleration.
The $\Psi_4$ Newman-Penrose scalar projection of the Riemann tensor is proportional to the second time derivative of the waveform~\cite{Cristofoli:2021vyo} and therefore also exhibits nontrivial dependence on $K$.


\Section{Effective Hamiltonian}
We construct a two-body Hamiltonian that describes the conservative evolution of the spinning bodies with nonvanishing $K$ vectors in general relativity. This Hamiltonian can inform EOB models for gravitational waveforms, connecting theoretical computations and experimental observations (see, e.g., Refs.~\cite{Buonanno:2024byg, Damour:2001tu, Damour:2008qf, Barausse:2011ys, Bini:2017xzy}).

The steps leading to the Hamiltonian are described in detail in Ref.~\cite{Bern:2023ity}, which builds on effective Hamiltonians used for the non-spinning~\cite{Cheung:2018wkq} and spinning cases without $\clK$~\cite{Bern:2020buy}. 
We start with an ansatz as the most general linear combination of spin structures $\Sigma_a$
(i.e., monomials in $\clS$,  $\clK$, coordinates and momenta up to the desired order in the former two~\footnote{Here $\clK$ is the operator form of the worldline $\text{K}$ in the rest frame.}) 
and determine their coefficients by matching the two-body scattering amplitudes with the field theory ones, including the products of external polarizations.
Thus, 
\begin{align}
\opH &= \sqrt{\opp ^2 + m_1^2}+\sqrt{\opp^2 + m_2^2} + V
\,, \\
V &= \sum_{a} V^{a} \spinStr{a}
\,, \quad
V^{a}=\sum_{n=1}^\infty 
    \left( \frac{G}{|\opr|} \right)^n c_n^{a}(\opp^2)
\,, \nn
\label{Eq:Hamiltonian_Definition}
\end{align}
where $m_1$ and $m_2$ are the masses of the two objects, and $\opr$ and $\opp$ are the relative distance and momentum in the COM frame.
In the text, we primarily focus on a single spinning body and suppress the particle label $\clS_1 \equiv \clS$ and $\clK_1 \equiv \clK$. The spin structures are listed in Table~\ref{tab:Operators}.
We collect the spin structures and their coefficients up to $\order(G^2)$ for both bodies spinning in the ancillary file {\tt Hamiltonian.m}.

\begin{table}
\setlength{\tabcolsep}{8pt}
	\begin{tabular}{c c c}
    \hline
    $\bm 1$ & 
    $\left(\opL \cdot \clS\right)/\opr^2$ &
    $\left(\opr \cdot \clK\right)/\opr^2$ \\
    $\left(\opr \cdot \clS \right)^2/\opr^4$ & 
    $\left(\opr \cdot \clK\right)\left( \opL \cdot \clS\right)/\opr^4$ &
    $\left(\opr \cdot \clK \right)^2/\opr^4$ \\
    $\clS^2/\opr^2$ &
    $\left(\clK \cdot \left( \opp \times \clS \right)\right)/\opr^2$ &
    $\clK^2/\opr^2$ \\
    $\left(\opp \cdot \clS \right)^2/\opr^2$ & 
    $\left(\opr \cdot \clS\right)\left( \left( \opr \times \clK \right) \cdot \opp\right)/\opr^4$ &
    $\left(\opp \cdot \clK \right)^2/\opr^2$ \\
    \hline
	\end{tabular} 
	\caption{Spin structures
     up to terms quadratic in $\clS$ and $\clK$ appearing in the Hamiltonian for a single spinning body.}
	\label{tab:Operators}
\end{table}

The resulting Hamiltonian coefficient for the structure
\begin{align}
\spinStr{(1,3)} = \frac{\opr \cdot \clK}{\opr^2} \,,
\end{align}
vanishes exactly, in agreement with the prediction of general relativity that the mass dipole of compact objects does not partake on its own in gravitational interactions. 
We expect this property to hold to all orders in Newton's constant and spin~\footnote{
When both particles are spinning, the Hamiltonian allows multiple interactions that are linear in the $K$ vector of one of the objects, e.g.,
$
\spinStr{(2,13)} = \left(\opr \cdot \clK_1\right)\left(\opL \cdot \clS_2\right)/\opr^4.$ The coefficients of all such structures vanish up to the order considered.} because general coordinate invariance guarantees that, to linear order in the spin tensor, the gravitational coupling of massive bodies is universal.

Our Hamiltonian, which governs the conservative evolution of our binary with additional degrees of freedom in general relativity, leads to spin-magnitude change.
Indeed, by solving Hamilton's equations, we find at $\order(G)$:
\begin{widetext}
\begin{align} 
\Delta \clS^2 = \frac{4G}{|\bm b|^2} \,\frac{m_2}{m_1} \left[
    2 D_2  \,
    \eftSigma 
    \left(S_{x}^{(0)}{\KxS}_{y}^{(0)}+S_{y}^{(0)}{\KxS}_{x}^{(0)}\right)
    +
    E_2 \, \frac{ 
    \left(2 \eftSigma^2 - 1\right) 
    }{ \sqrt{\eftSigma^2-1}} \left(K_{y}^{(0)}{\KxS}_{y}^{(0)}-K_{x}^{(0)}{\KxS}_{x}^{(0)}\right) \right]  + \order(G^2) \, ,
\end{align}
\end{widetext}
where $\clS^{(0)}$ and $\clK^{(0)}$ are, respectively, the initial values of the rest-frame $S$ and $K$ vectors, and $\clKxS^{(0)}\equiv\clK^{(0)}\times\clS^{(0)}$. 
In our COM frame, the impact parameter $\bm b$ points in the $x$ direction, the incoming 3-momenta $\bm p_1$ and $\bm p_2$ are along the $z$ direction,
and $\eftSigma= {p_1\cdot p_2}/{m_1m_2}$, as described in detail in  Ref.~\cite{Bern:2023ity}. As noted there, while neither $\clS^2$ nor $\clK^2$ are conserved, the difference $\clS^2 - \clK^2$ is, so we have $\Delta \clK^2 = \Delta \clS^2$.
We give the observables obtained by solving Hamilton's equations in 
the ancillary file {\tt Observables.m}.
For $\clK^{(0)}=0$ or for $D_2=E_2=0$ we reproduce the impulses and spin kicks obtained in Refs.~\cite{Bern:2020buy,Kosmopoulos:2021zoq}, which match those obtained from worldline theories with an SSC imposed~\cite{Liu:2021zxr,Jakobsen:2021zvh}~\footnote{At this order, either $\clK^{(0)}=0$ or $D_2=E_2=0$ is sufficient for this purpose. We expect that {\em both} conditions are necessary at higher orders.}.


\Section{Eikonal Phase} 
The spin-dependent scattering observables are encoded in the eikonal phase~\cite{Bern:2020buy,  Gatica:2023iws, Luna:2023uwd}, which to $\order(G^2)$ is given by the two-dimensional Fourier transform (from $\bs q$ space to $\bs b$ space) of the classical part of the two-body amplitude found in the ancillary files.  Ref.~\cite{Bern:2023ity} gave a generalization of the construction of Ref.~\cite{Bern:2020buy} that includes the effects of $K$ and applied it to electrodynamics.  Here, we confirm the validity of this construction through $\order(G^2 \sS^2)$ in gravity.
From Ref.~\cite{Bern:2023ity}, the transverse impulse~\footnote{Here $\clp\cdot\bs b=0$, and so all the $\bs b$-derivatives are projected orthogonal to the incoming momentum $\clp$. The contributions along $\clp$ are instead obtained using energy conservation.  See Ref.~\cite{Bern:2023ity}.
}, 
and the $S$ and $K$ changes are given by
\begin{align}
    \Delta\mathbb{O} &= \{\chi,\mathbb{O}\} + \frac{1}{2}\big\{\chi,\{\chi,\mathbb{O}\}\big\} \nonumber\\
    & \hskip 1 cm  + \mathcal{D}_L\big(\chi,\{\chi,\mathbb{O}\}\big) - \frac{1}{2}\big\{\mathcal{D}_L(\chi,\chi),\mathbb{O}\big\} \,,
\label{eqDeltaO}
\end{align}
where $\mathbb{O}=(\clp_\perp,\mathbf{S},\mathbf{K})$ 
and we define $\{f,\clp_{\perp}\}\equiv\frac{\doe f}{\doe \bm b}$,
\begin{align}
\label{defineDL}
\mathcal D_L(f,g)&\equiv -\epsilon_{ijk}\bigg(S_{i}\frac{\doe f}{\doe S_{j}}
  + K_{i}\frac{\doe f}{\doe K_{j}}\bigg)\frac{\doe g}{\doe L_k}\, ,
\end{align}
with an obvious generalization for two spinning particles. The Lorentz algebra gives the brackets of $\clS$ and $\clK$ (see Eq.~(5.40) of Ref.~\cite{Bern:2023ity}). We have verified that Eq.~\eqref{eqDeltaO} reproduces the impulse, $S$ and $K$ change from Hamilton's equations for a single spinning body. 


\Section{Conclusions and Outlook}
%
The body-fixed tetrad has six degrees of freedom. Demanding the minimal EFT description of the body causes only three of them to be dynamical and requires an SSC to fix the other three. 
Relaxing minimality removes the need for an SSC and promotes the additional tetrad components to new dynamical variables $K$.

We demonstrated that the traditional description of gravitating spinning compact objects can be naturally extended with such additional degrees of freedom.
The $K$ vector yields new interaction terms with associated free Wilson coefficients in the field-theory, worldline, and two-body-Hamiltonian descriptions, which are all in physical agreement. This mirrors the conclusions of Ref.~\cite{Bern:2023ity} obtained in electrodynamics.

In agreement with the principles of general relativity, interactions linear in $K$ at $\order(\sS)$ do not affect physical observables. 
We explicitly demonstrated in general relativity that multilinears in $K$ and $S$ not only affect the impulse and spin kick at $\order(G)$ and $\order(G^2)$ but also modify the waveform in ways that cannot be replicated by merely readjusting the Wilson coefficients at vanishing $K$.
We also explicitly showed that a $K$-dependent generalization~\cite{Bern:2023ity} of the spinning eikonal formula~\cite{Bern:2020buy} yields the correct impulse, $\Delta S$ and $\Delta K$. It would be interesting to see whether this agreement continues to higher orders.

It remains an interesting question whether a nonvanishing $K$ vector leads to measurable effects for any compact astrophysical bodies in our Universe~\vphantom{\cite{Kuntz:2021ohi, Kuntz:2022onu}}\footnote{See Refs.~\cite{Kuntz:2021ohi, Kuntz:2022onu} which discuss the hierarchical three-body system in which the inner body may be interpreted as exhibiting a nonzero $K$ vector.}. The Wilson coefficients governing its interactions could be determined by matching them onto suitable models or comparing theoretical waveforms to numerical-relativity simulations or gravitational-wave data.
While we focused here on the scattering regime, the absence of the tail effect below ${\cal O}(G^4)$ makes it straightforward to use our Hamiltonian as well as its extensions through ${\cal O}(G^3)$ and to higher powers of $\clS$ and $\clK$ in the bound regime. 
It would be very interesting to contrast the $\clK\ne 0$ and $\clK=0$  bound-orbit waveforms; we expect that, similarly to their scattering counterparts, $\clK$ will have observable effects. 
It would also be interesting to explore, both in a scattering and bound setting, the degeneracy between $\clK=0$ and aspects of physics beyond general relativity.  

In this work, we observed, to low orders, that an SSC may be imposed {\em a posteriori}, by choosing Wilson coefficients so that all $K$-dependent contributions cancel out. The results are compatible with traditional descriptions of black holes and neutron stars. An all-order proof of this property is desirable.
In both field theory and worldline formalisms, an unconstrained description of spin, such as the one advocated here, can facilitate calculations at higher orders in both spin and Newton's constant.

\Section{Acknowledgments}
We thank Juan Pablo Gatica, Donal O'Connell and Mikhail Solon for useful discussions.  
We especially thank Justin Vines for many important and valuable discussions and for collaborating on earlier work~\cite{Bern:2023ity}, which is the basis for much of this paper.
Z.~B. and T.~S.  are supported by the U.S. Department of Energy
(DOE) under award number DE-SC0009937. 
D.~K. is supported by the Swiss National Science Foundation under grant no. 200021-205016.
A.~L. is supported by funds from the European Union’s Horizon 2020 research and innovation program under the Marie Sklodowska-Curie grant agreement No.~847523 ‘INTERACTIONS’.  
R.~R. and F.~T.~are supported by
the U.S.  Department of Energy (DOE) under award number~DE-SC00019066.
We also thank the Mani L. Bhaumik Institute for Theoretical Physics for support.


\bibliography{ref}{}
\bibliographystyle{apsrev4-1}

\end{document}